# MontiCore: a framework for compositional development of domain specific languages

**Holger Krahn · Bernhard Rumpe · Steven Völkel**

**Abstract** Domain specific languages (DSLs) are increasingly used today. Coping with complex language definitions, evolving them in a structured way, and ensuring their error freeness are the main challenges of DSL design and implementation. The use of modular language definitions and composition operators are therefore inevitable in the independent development of language components. In this article, we discuss these arising issues by describing a framework for the compositional development of textual DSLs and their supporting tools. We use a redundance-free definition of a readable concrete syntax and a comprehensible abstract syntax as both representations significantly overlap in their structure. For enhancing the usability of the abstract syntax, we added concepts like associations and inheritance to a grammar-based definition in order to build up arbitrary graphs (as known from metamodeling). Two modularity concepts, grammar inheritance and embedding, are discussed. They permit compositional language definition and thus simplify the extension of languages based on already existing ones. We demonstrate that compositional engineering of new languages is a useful concept when project-individual DSLs with appropriate tool support are defined.

H. Krahn · B. Rumpe · S. Völkel (✉)
Software Engineering Group, Department of Computer Science 3,
RWTH Aachen University, Aachen, Germany
e-mail: voelkel@se-rwth.de
URL: http://www.se-rwth.de

H. Krahn
e-mail: krahn@se-rwth.de

B. Rumpe
e-mail: rumpe@se-rwth.de

## 1 Introduction

Software development is a complex task which involves different activities. To push the border of development further and reduce the costs and risks of complex software development, many actions are necessary. One of them is the use of domain specific languages (DSLs) which are generally languages that specifically fit the domain the software under design is assisting [9].

As the experience with DSLs grows and with them the demand to capture domain-specific concepts, DSLs become increasingly complex. This makes them significantly harder to evolve, ensure error freeness, etc. DSLs therefore themselves become a target of management, evolution, and in particular reuse. This is especially important in situations where grammars and languages change continuously [3]. From programming languages we know that modularity of the units described, and a semantically clear and precisely understood composition of the modules is a key technique to handle this kind of complexity [55].

An appropriate modularity concept for DSLs and corresponding composition operations that permit independently developed language parts to be composed are therefore inevitable. As discussed in [29] this does not only include the syntactical aspects of composition but also its semantics (in terms of meaning [26]), when to use it methodologically, and how context conditions such as typing information, etc. fit together. This paper shows how these issues are addressed in the framework Monticore (e.g., [38,39,41,49]).

The development of a new language incorporates different activities. A concrete syntax and an abstract syntax is developed. Sometimes a more or less explicit and formal semantics is defined for the language [26]. These activities are complemented by developing a type system, priorities for operators, naming systems, etc., if appropriate. When examining this process the definition of concrete and abstract syntax show a significant redundancy. This leads to duplications and therefore possible inconsistencies, which are a constant source of problems in an iterative agile development of languages within model-driven development. The situation becomes far worse when modular development of the language is desired. Then every language module comes equipped with concrete and abstract syntax. All parts and all syntax versions need to evolve in parallel. The synchronized co-evolution in this situation unnecessarily complicates an efficient development and evolution of languages. Therefore, an integrated development of both is highly desirable and was chosen in our approach.

As we did not want to concentrate on graphical tooling issues, but wanted an efficient way to develop models as well as tools, we have chosen a textual language as the frontend. This allows us to reuse ordinary textual editors (augmented with syntax highlighting) as known from Eclipse [18,24] as well as parser-generators like Antlr [57] or SableCC [17] to generate language recognition tools from this form of language definition.

While graphical modeling approaches are popular because they provide an easy overview and therefore access to the model, our experience is that textually defined models are a lot more efficient to handle and manipulate [24]. So textual DSLs do have advantages for the modeler as well as for the language designer. The latter has an easier task to handle and can reuse well-understood tools including version control and diffs. A sole disadvantage for the modeler, however, is that graphical languages are usually more intuitive to read.

Beside grammar-based approaches, metamodeling is a popular method to define the abstract syntax of modeling languages. We added elements like associations and inheritance between the abstract syntax tree nodes to our grammar format. Thus, the instance graphs are full-fledged graph structures with a spanning tree in it. This permits both, arbitrary graph handling mechanisms and tree-based navigation to be used at will.

MontiCore can be used for the agile development of simple as well as complex textual languages. These may be specific versions of programming languages, logic, textual representations of graphical modeling languages, or any other form of domain-specific languages. MontiCore provides a grammar format in such a way, that the recognition power of the resulting parser is only limited by the underlying parser generator, namely Antlr [57], which is a predicated-LL(k) parser generator. Furthermore, the framework provides means of compositionality and modularity on the language level.

The MontiCore framework is also delivered as an Eclipse plugin including an editor with different comfort functionalities like syntaxhighligthing, outlines, and code-completion. Furthermore, it analyzes the input [esp. the grammar and its LL(k) property] and reports errors and warnings using Eclipse problem reports. However, these functionalities are built on top of the core framework which is therefore usable on the command line as well. This is especially important for build scripts.

As key results of the work described here, we discuss language embedding and inheritance. Inheritance allows a developer to apply incremental changes to a language. Compositional embedding is useful to combine different language fragments to a new coherent one. Because modularity concepts are not only highly desirable for the concrete syntax but also for other artifacts (abstract syntax, tools, etc.) under design, the MontiCore framework provides a coherent concept of modularity for different aspects of the language.

This article is based on earlier work [22,23,25,38–41,49]. In particular, a previous version of the work described here was already described in [39,41]. This article was enhanced and extended to reflect all results found when exploring MontiCore's capabilities with respect to compositional language definition.

Section 2 describes the syntax of the MontiCore grammar format and its semantics in form of the resulting concrete and abstract syntax of the defined language. Section 3 explains the different modularity concepts in the MontiCore framework and what can be achieved with their use. Section 4 explains how further concepts can be used to build domain-specific tools based on the language definition in a modular fashion. Finally, Sect. 5 relates our approach to others and Sect. 6 concludes the paper.

## 2 Language definition using MontiCore

One core element of the MontiCore framework is the grammar definition language that allows defining a concrete textual syntax as well as the internal representation (abstract syntax) of a language. For this purpose, we use an enriched grammar that we will introduce in the following.

To represent both elements of a language definition in a single document has already been discussed intensively in compiler research. We decided to use a single-definition approach in contrast to others like [33] to simplify the development of DSLs. The benefits are a single concise language definition developers and users can rely on. Problems like keeping two descriptions consistent cannot occur. In addition, the abstract syntax matches the concrete syntax of the language automatically in the sense that similar elements are represented similar and distinct elements are represented differently. To our experience this design guideline [30] is often neglected by unskilled language developers which is assured in our approach by construction.

Of course, there are some counter-argument against an integrated definition of both syntaxes. First, languages may

```
                    ──────── MontiCore-Grammar ────────
1  ShopSystem =
2    name:IDENT (Client | PremiumClient)*
3    (OrderCreditcard | OrderCash)*;
4
5  Client =
6    "client" name:IDENT Address;
7
8  PremiumClient =
9    "premiumclient" name:IDENT discount:NUMBER Address;
10
11 OrderCreditcard =
12   "creditorder" clientName:IDENT billingID:NUMBER;
13
14 OrderCash =
15   "cashorder" clientName:IDENT amount:NUMBER;
16
17 Address =
18   street:STRING town:STRING;
```

**Fig. 1** Defining productions in MontiCore

```
                    ──────── MontiCore-Grammar ────────
1  // Simple name
2  token IDENT = ('a'..'z'|'A'..'Z')+ ;
3
4  // Numbers (using default transformation)
5  token NUMBER = ('0'..'9')+ : int;
6
7  // Cardinality (STAR = -1)
8  token CARDINALITY = ('0'..'9')+ | '*' :
9    x -> int {
10     if (x.equals("*")) return -1;
11     else return Integer.parseInt(x);
12   };
```

**Fig. 2** Definition of token classes in MontiCore

have different concrete but only one abstract syntax. From our experiences, this argument can mostly be neglected for DSLs as often the (only) concrete syntax emerges the domain. The second counter-argument is that abstract syntax is often different from concrete syntax so that tasks of semantic analysis can focus on the structure of the language without being overrun by syntactical issues. While this is generally true, we made the experience that—in contrast to GPLs—DSLs often have simple type systems and context conditions which can be implemented straightforward on top of your automatically provided abstract syntax. However, we will describe means provided by MontiCore which can be used in order to influence the abstract syntax without affecting the concrete syntax. These means can be used to gain a more desirable and comfortable version while still guaranteeing consistency between both syntaxes. Nevertheless, if it is really necessary to have a different abstract syntax—either for comfort reasons or to separate both because of multiple concrete syntaxes—one can define this abstract syntax and use a model transformations to transform them into another which is a well-accepted approach in the model-driven community.

### 2.1 Defining concrete and abstract syntax

The MontiCore grammar format is an enriched context-free grammar that is derived from the input language of common parser generators. Figure 1 contains a simple example demonstrating some core definitions.

A production has a name and body (right hand side) separated by "=". The body contains nonterminals, token classes, and terminals. The usual concepts for structuring the body are alternatives (separated by "|"), blocks (in parenthesis) and repetition by adding cardinalities. Blocks, nonterminals, token classes, and terminals can have cardinality "?" (optional), "*" (unbounded cardinality) or "+" (at least one). Furthermore, nonterminals, token classes, and terminals can be named (in form of a prefix like "name:") in order to access these elements in the internal representation of the abstract syntax.

Token classes are usually handled as strings, but more complex data types are possible by giving a function defined in the programming language Java that maps a string to an arbitrary data type. Default functions exist for primitive data types like floats and integers. Figure 2 embodies some illustrating examples for token classes. Line 2 defines IDENT being mapped to string. NUMBER in line 5 is mapped to an integer as provided by the default mapping. In line 8 CARDINALITY is mapped to an int. However, we use a special mapping that is defined by the Java code below (line 10 and 11) where the unbounded cardinality is expressed as the value −1. To simplify the development of DSLs, the token classes IDENT and STRING are predefined to parse names and strings.

In addition to the already explained token classes, terminals like keywords can be added to the concrete syntax of the language. These elements are normally not directly reflected in the abstract syntax. Note that in contrast to many parser generators there is no specific need for distinguishing between keywords like "public" and special symbols like ",". To further simplify the development of a language we generate the lexer automatically from the grammar. For this purpose we generate the Antlr-literals for all terminals of the grammar, generate standard lexical symbols for identifiers and strings, etc.[1] By this strategy a number of technical details like the distinction between parser and lexer (necessary for the parser generator) are hidden from the language developer as far as possible.

The abstract syntax (also known as the internal representation) of a language is also derived from this grammar.

---

[1] The generation of standard lexical symbols can be switched off by special options in the grammar specification. This is sometimes necessary if the language uses different kinds of identifiers whereas in most cases the predefined symbols are adequate.

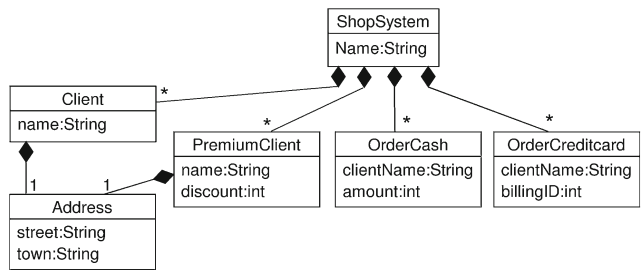

**Fig. 3** Abstract syntax derived from the grammar shown in Fig. 1 as UML class diagram [52]

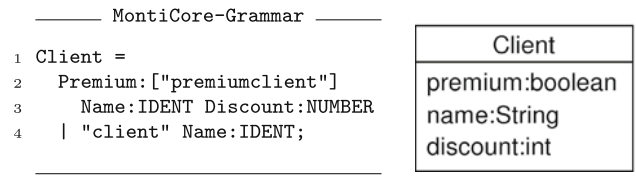

**Fig. 4** Use of constants

In Fig. 3 we see the class definitions derived from the grammar in Fig. 1. Each production of the grammar leads to a class with the same name in the abstract syntax.

The body of the production determines the attributes of the class as follows:

Nonterminals. Each nonterminal is mapped to a composition of the corresponding type/class.
Explicitly named elements. Nonterminals, token classes and terminals can explicitly be named to determine the attribute or composition name, where it is stored. For example, `street:STRING` is directly mapped into an attribute street of type String.
Optional elements. Optional elements like `Client?` are still mapped to ordinary attributes but such an attribute may be `null`.
Repeated elements. An element, for example nonterminals, may be repeated in two ways. In the simplest form, it can be marked with the Kleene star. The second possibility is that there exists a derivation for a production where this element occurs at least twice. The term `A ( B | A C )?` for example can be derived to `A A C` with 2 A's. Both cases lead to a cardinality higher that one and instead of an ordinary attribute, a list is used in the abstract syntax. This approach allows specifying constant separated lists without an extra construct in the grammar format. Thus, the term `A:X (","  A:X)*` results in an unbounded composition `A` with minimal cardinality of one that contains all occurrences of `X`. Please note that both occurrences use the same referenced rule and name. If such a unification is undesired because the order of the appearance should be reflected, different attribute names have to be used.
Handling alternatives. In the abstract syntax, we flatten alternatives by representing each alternative, e.g., of `B|C` by an individual attribute. Both attributes `b:B` and `c:C` are given, but only one will actually have a value, whereas the other will be `null`.
Handling blocks. Blocks are merely used to structure the concrete syntax, such as in `(B|C)*`. Like with handling alternatives, blocks are flattened in the abstract syntax representation. The above `(B|C)*` leads to two lists of `B` and `C` objects.
Constants and symbols. Keywords are normally not reflected in the abstract syntax. If they are optional and their appearance changes the meaning of the model, their presence can be added to the abstract by using them in brackets. In Fig. 4 the reserved word `premiumclient` determines the value of the attribute `premium`. A single terminal inside brackets is translated to a boolean attribute as shown, and a list of constants (separated by `|`) is mapped to an enumeration attribute.

Please note that when deriving the abstract syntax from the grammar we so far have made two not quite straightforward decisions. First, whereas flat grammar rules and object-oriented realizations of the abstract syntax widely coincide, there is also a structural mismatch, caused by alternatives within blocks [44]. While `A=B|C` could be handled using subclasses `B` and `C` of `A`, a definition like `A = (B|C) D` cannot be reflected directly at all. Our approach is to provide a relatively flat abstract syntax representing each nonterminal `B` and `C` as attribute in class `A`. An alternative is to restructure not only the grammar to `A=X D` and `X=B|C` which leads to a better structure of the abstract syntax classes, but also to more classes.

Second, we also allow using blocks within the grammar and flatten those in the abstract syntax. Like above, this leads to loss of potentially important information and the developer should not nest blocks too deeply. In the above example, the exact order of appearance of the `B`s and `C`s in `(B|C)*` cannot be reconstructed from the abstract syntax only. However, to solve this problem each class in the abstract syntax is associated with a source position in the text. Using this information, the order can be reconstructed easily if necessary.

The MontiCore mechanism to define concrete and abstract syntax is rather expressive and comfortable. However, it needs to be used carefully to avoid oversimplification and loss of necessary information, especially when combining name derivation, blocks, and alternatives.

As already stated above, MontiCore is based on Antlr and thus uses a predicated-LL(k) parsing mechanism. There are two main restrictions for LL(k) parsers: first, left-recursive grammars cannot be used. This can be solved by transforming the grammar into a right-recursive version. The drawback is, however, that this influences the abstract syntax. Alternatively, keywords can be inserted what preserves the abstract syntax. Second, grammars that are not LL(k) for any k cause problems in standard-LL(k)-parsers. As ANTLR uses predicated-LL(k), this does not hold as syntactic and semantic predicates [56] can be used in this case. The main reason for choosing ANTLR as underlying parser generator was its maturity and good documentation. The recursive-descent style of the generated parsers allowed us to easily instrument the generated code in order to create our specific AST-structure and our enhancements to modularize languages.

Figure 3 shows a UML class diagram of the abstract syntax that is created from the productions. In the MontiCore framework this class diagram is mapped to a Java implementation. All attributes are realized as private fields with appropriate access methods (get/set). Composition relationships are realized as attributes and contribute to the constructor parameters of the class. Unnamed compositions use the name of the opposite class for the access methods. To handle the abstract syntax, the MontiCore framework provides an infrastructure to handle the abstract syntax. For example, all classes support a variant of the Visitor pattern [19] to traverse the abstract syntax along the composition relationships.

Both versions of our `ShopSystem` language do have some deficiencies. The version defined in Fig. 1 suffers from the problem that both productions `Client` and `PremiumClient` as well as `OrderCash`/`OrderCreditcard` share some common substructure, but are not related at all. In a second version, we replace the substructure `Client|PremiumClient` by `Client` as shown in Fig. 4. Then an additional invariant is needed in the abstract syntax that is not visible from the class structure, namely only if the boolean flag `premium` is true, the `discount` may be defined.

These deficiencies motivate the extension of the MontiCore language definition format to include more advanced features that handle often occurring challenges. Therefore, we use object-oriented features like inheritance, interfaces, and associations in the abstract syntax.

## 2.2 Interfaces and inheritance between nonterminals

The abstract syntax shown in Fig. 1 raises the question if something like an "interface-nonterminal" `Order` could be defined that is realized by the ordinary nonterminals `OrderCreditcard` and `OrderCash`. In a traditional attempt, we would use `Order=(OrderCreditcard | OrderCash)` or if a common part X can be factored out: `Order = X (OrderCreditcard|OrderCash)`. The extensions `OrderCreditcard` and `OrderCash` contain the variants that `Order` has. However, this approach has two drawbacks:

First, the common part X need to be at the beginning or at the end of the production, which may not be feasible in the concrete syntax at all. This approach does not work with `Order` in the shown example. Second, this approach lacks extensibility. The definition of top nonterminal `Order` does know what its alternatives are. The language is fully defined and cannot be extended. This is in strong contrast to object-oriented concepts, where the superclass is being defined without knowing what its alternatives (subclasses) are. Therefore, we prefer `B extends A` instead of `A = B|...`[2].

For this purpose we extend the MontiCore grammar language by a concept expressing an inheritance relationship between nonterminals and a concept of an interface-nonterminal that can be implemented by nonterminals.

### 2.2.1 Inheritance of nonterminals

A nonterminal inherits from a super-nonterminal using the keyword `extends` as shown in Fig. 5. In line 16 `PremiumClient` extends a given nonterminal `Client`. Inheritance between nonterminals has two consequences. First, we translate the inheritance relationship in the grammar to an object-oriented inheritance relationship between the according classes in the abstract syntax. In addition, we generate only those attributes in the subclass which were not already defined for the superclass. Thus, `PremiumClient` does not have an attribute `Name` because this is already part of its superclass `Client`. Second, this extension also modifies the concrete syntax and therefore the parser. Inheritance adds an additional alternative to the super-production, just like `Client=...|PremiumClient`.

The EBNF section in Fig. 5 shows a representation with equivalent concrete syntax to explain the mapping of the MontiCore grammar format to the input format of a parser generator.

As said, this concept is motivated by the definition of object-oriented inheritance where each occurrence of a superclass can be substituted by a subclass object. We have decided to use this object-oriented style of inheritance instead of the traditional grammar style to be more flexible when extending languages. In the left grammar, the production `Client` needs not be changed when extending the language with `PremiumClients`. This is a significant benefit that we will further explore when defining inheritance on languages in Sect. 3.1. As a disadvantage, it becomes more complicated to understand the language as several places need to be looked up to understand the variants of an extended nonterminal like `Client`. For this purpose, we automatically generate an EBNF version of the grammar which resolves all extending rules as shown in Fig. 5, line 15 on the right.

---

[2] This is especially important for grammar inheritance as we will see in Sect. 3.1. It may be the case that another nonterminal that extends `Order` is defined in a subgrammar. Changing the supergrammar in that case is not desirable since the supergrammar then depends on the subgrammar and cannot be used without it.

```
                    MontiCore-Grammar
 1  ShopSystem =
 2    Name:IDENT
 3    Client* Order*;
 4
 5  OrderCreditcard implements Order =
 6    "creditorder"
 7    ClientName:IDENT BillingID:NUMBER;
 8
 9  OrderCash implements Order =
10    "cashorder"
11    ClientName:IDENT Amount:NUMBER;
12
13  Client =
14    "client" Name:IDENT Address;
15
16  PremiumClient extends Client =
17    "premiumclient"
18    Name:IDENT Discount:NUMBER;
19
20  Address =
21    Street:STRING Town:STRING;
22
23  interface Order;
24
25  ast Order =
26    ClientName:IDENT;
```

```
                         EBNF
 1  ShopSystem ::=
 2    IDENT Address
 3    Client* Order*
 4
 5  OrderCreditcard ::=
 6    "creditorder"
 7    IDENT NUMBER
 8
 9  OrderCash ::=
10    "cashorder"
11    IDENT NUMBER
12
13  Client ::=
14    "client" IDENT Address
15    | PremiumClient ;
16
17  PremiumClient ::=
18    "premiumclient"
19    IDENT NUMBER
20
21  Address ::=
22    STRING STRING
23
24  Order ::=
25    OrderCredit | OrderCash
26
```

**Fig. 5** Inheritance and use of interfaces

Sometimes we found it useful to only extend the abstract syntax through inheritance, without affecting the concrete syntax. For this purpose, keyword `astextends` can be used to express an inheritance that is restricted to the abstract syntax and does not influence the concrete syntax.

### 2.2.2 Interfaces between nonterminals

The form of inheritance introduced above also allows the definition of interfaces like `Order` that are implemented by ordinary nonterminals. In Fig. 5, line 23 left, the nonterminal `Order` is introduced as an interface with no concrete syntax. `Order` acts as ordinary nonterminal, for example, like shown in line 3. The keyword `implements` is used to implement an interface (Fig. 5, line 5 and 9, left) with the effect that the nonterminal `Order` is defined as an alternative production (see Fig. 5, line 24 right).

An analogous keyword called `astimplements` combines only the abstract syntax in the same way as explained for superclasses. Interface-nonterminals can be defined like normal rules using the additional keyword `interface` and may also extend other interfaces thus enabling the full power of object-oriented mechanisms. If the rule body is left empty like shown in the example (Fig. 5, line 23, left) all implementing rules separated by | form the default body of this rule.

To complete the availability of object-oriented concepts in the MontiCore grammar, we have added the concept of an abstract nonterminal. The keyword `abstract` can be used to define an abstract class in the same way as interfaces are defined. As we will later see, the behavior can be specified in form of Java methods within a class generated from a nonterminal. In conformance to Java, abstract classes allow to specify behavior in the class which is inherited to all subclasses, whereas interfaces do not have behavior.

By default interfaces and abstract classes do not contain attributes. We decided against an automatic strategy where all common attributes of known subclasses are extracted, as interfaces typically are good places for future extensions of the defined language which may only use a subset of all available attributes. Additional attributes may be added to interfaces and classes by using the keyword `ast` like shown in the example (Fig. 5, line 25, left). This concept uses the same syntax as in an normal production, but only adds attributes to the abstract syntax and does not affect the concrete syntax. The attributes of interfaces are realized as get- and set-methods in the implementation and can therefore be used in the Java implementation (as Java interfaces cannot contain fields but only methods).

For a clarification of the resulting data structure, the abstract syntax resulting from the language definition in Fig. 5 is shown in Fig. 6.

### 2.3 Associations

The attributes `Name` in `Client` and `ClientName` in `Order` (see Fig. 6) are obviously semantically connected. However, context-free syntax definitions cannot capture

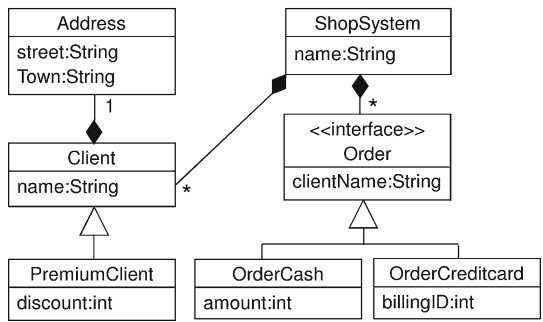

**Fig. 6** Abstract syntax of the language defined in Fig. 5

```
                                    MontiCore-Grammar
 1  OrderCreditcard implements Order =
 2    "creditorder" ID:IDENT Amount:IDENT;
 3
 4  OrderCash implements Order =
 5    "cashorder" ID:IDENT Amount:IDENT;
 6
 7  Client =
 8    "client" Name:IDENT Address;
 9
10  PremiumClient extends Client =
11    "premiumclient" Name:IDENT Discount:IDENT;
12
13  association ClientOrder
14    Client 1 <-> * Order.orderingClient;
```

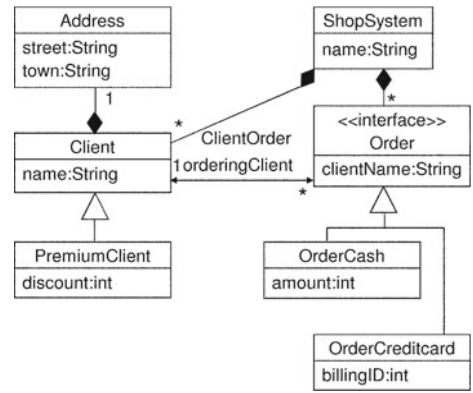

**Fig. 7** Specification of associations

these connections adequately. In the example we are interested in establishing an invariant that an `Order` may only use `Client` names that exists. Furthermore, an efficient navigation from the usage of such a name to its definition to look up additional information is necessary.

When designing a meta-model, this relation is usually expressed by an association where an order references a client as the ordering person. However, these associations do break the tree structure that a grammar produces and create an ordinary graph. The MontiCore language definition extends the context-free grammar by adding a mechanism of defining associations like these. The result of this extension is an arbitrary graph with an embedded spanning tree of compositions that results from the original grammar.

In a language definition the keyword `association` allows specifying non-compositional associations between rules which enables the navigation between objects in the abstract syntax. With this mechanism we can define uni- and bidirectional navigation between objects of the abstract syntax.

An example for an association can be found in Fig. 7 (line 13–15) where the association `OrderingClient` connects one `Order` object with a single `Client`. As associations are implemented via attributes, the reverse direction is a second attribute that is named `Order` which connects one `Client` object with an unbounded number of `Order` objects. This form is very similar to the associations in EMF [7] where two associations are specified separately but are related to each other via the attribute "isOppositeOf".

The main challenging question for associations in a unified format for concrete and abstract syntax is not the specification of the associations, but the automatic establishment of all links between associated objects in a step after parsing. Grammar-based systems usually parse the linear character stream and represent it in a tree structure in accordance to the grammar. All additional connections necessary are established through definitions and usage of names (of classes, methods, attributes, states, objects, etc.). Symbol tables are calculated and later used to navigate between nodes in the abstract syntax tree (AST). The desired target of navigation is determined by identifiers in source and target nodes and a name resolution algorithm.

Due to the simple nature of many domain-specific languages that lack complex namespaces, simple resolution mechanisms like file-wide unique identifiers can often be used for creating links. Of course, this simplification is not always suitable, for example languages like Java and many UML-sublanguages use a more sophisticated namespace concept. In order to integrate support for such complex languages in a language definition framework like MontiCore, the scoping and resolution mechanism have to be formalized in way that a developer can configure them for the language under design in a simple way. On the contrary, complex languages that use inheritance as a language concept and support models that are distributed among multiple files have complex and widely varying requirements. For example, the Java Language Specification [20] (certainly more complex than common DSLs) describes the name resolution algorithm on 14 pages and access control on another 13 pages in natural language. Especially static imports, inner classes, and inheritance complicate the problem in such a way, that this resolution mechanism seems to be inappropriate to be reused for another language without major changes.

Therefore, we use a twofold strategy: first, we generate interfaces that contain methods induced by the association to navigate between the AST-objects. The resulting classes of the abstract syntax allow accessing the associations in the same way as attributes and compositions are accessed. Second, we generate default implementations for simple resolving problems like file-wide flat simple or simple hierarchical namespaces. As an alternative for the second step, the DSL developer can instead program his own resolution algorithms if needed. Thus, complex and difficult formalizations of scoping definitions, different inheritance possibilities, and name resolution algorithms are avoided and replaced by a programming interface. As a second alternative the developer can extend the MontiCore framework by adding new forms of name analysis.

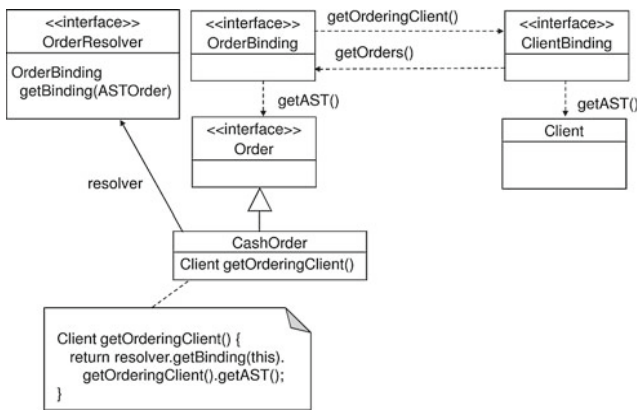

**Fig. 8** Java implementation of an association

Figure 7 extends the example from Fig. 6 by adding an association definition. The association `ClientOrder` connects each `Order` to a single `Client` (as specified by "1") and each `Client` to an unbounded number of `Orders` (specified by "*"). Please note that if the name of the association end is omitted, its realizing attribute is named after the target class in de-capitalized form. In addition to the shown cardinalities, ranges like "3..4" are possible values.

The links are established after parsing is completed. This way, links are not immediately stored upon node creation. The realization is designed in this way to easily support forward references in languages, which means that identifiers refer to elements that occur at a later position in the text file.

Figure 8 sketches the structure of a Java implementation for the class diagram from Fig. 7 with the most important methods. A `Binding`-interface is generated for each interface and class that is involved in an association as either source or target. This `Binding`-interface contains the relevant methods signatures for the navigation between different nodes. In addition, a `Resolver` is generated for each class or interface which can be used to effectively navigate between AST-objects while the actual navigation mechanism and its calculation are effectively hidden from the language developer.

Note that these interfaces are generated to simplify the establishment and use of associations for a DSL. If standard resolving algorithms are appropriate, MontiCore can generate both `Binding`-implementations and a single `Resolver`-implementation that resolves all objects automatically and therefore allows for an effective navigation between nodes. The complexity of multiple classes with different responsibilities is hidden from the user of the abstract syntax, for example, a programmer of a code generator for the developed language. He simply uses the get- and set-methods like `getOrderingClient` that returns the appropriate `Client` object.

From the point of view of the developer there is no difference between links established by parsing and links established by the association. This makes our generated abstract syntax comparable to metamodeling approaches where related objects are linked directly and not because of naming schemes. Furthermore, developers simply use the generated methods in order to access connected objects and do not have to care about the way the objects have been linked.

## 3 Modularity concepts

In [34] the term *grammarware* is coined as collective term for grammars and grammar-dependent software. MontiCore is categorized as a *meta-grammarware* that uses an enriched grammar format "as an executable specification (or a program)" to generate components as described above. Modularity principles for the language definition help to break down the complexity of a problem into smaller pieces and to increase reusability of "language modules". Each such language module shall be understandable by itself without having to consider internal knowledge about other pieces.

MontiCore supports two modularity concepts which can be used for different purposes. First, grammars may inherit from each other to add new productions or to override existing ones in order to adopt an existing language to new needs. Second, language embedding can be used in order to combine separately designed languages or parts of it. It is important to notify that these modularity concepts at first apply to the languages concrete syntax, but they also apply to the AST (internal representation), to the check of context conditions, to analysis and code generation techniques, to an independent development and composition of the tool infrastructure for language parts, and finally to a methodical decoupling of the language development and use. In Sect. 4 we will explain how MontiCore supports a modular development beyond concrete and abstract syntax.

```
                  MontiCore-Grammar
1 package mc.javasql;
2
3 grammar JavaSQL extends
4   mc.java.JavaDSL, mc.sql.SQLDSL {
5
6   SQLSelect implements Expression;
7 }
```

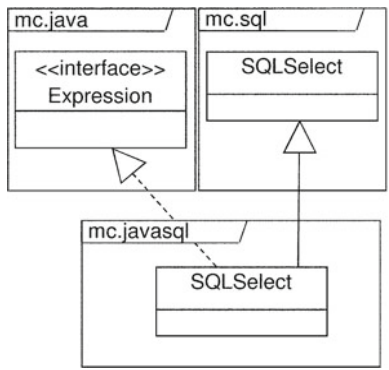

**Fig. 9** Multiple language inheritance

3.1 Grammar inheritance

Grammar inheritance can be used if an existing language shall be extended by specifying only the differences between a given and the new language. The existing language definition remains unchanged. A well-known example for such language extensions is the use of SQL-Select statements as expressions inside a general purpose language (GPL) (like prominently shown in [46]). The language can be extended by using a given grammar for the GPL and adding the SQL part only. In a monolithic approach, the grammars would be integrated and a new parser would be generated from there. This is obviously not desirable; a reuse of existing artifacts should be preferred.

MontiCore provides the concept of grammar inheritance that allows reusing an already existing infrastructure for the SQL language. It is possible to conveniently define the new grammar by inheriting from both, the GPL and the SQL grammar. Therefore, grammar inheritance in MontiCore allows a developer to specify (multiple) grammars from which all productions are inherited to the new grammar. This way, we achieve reuse of independently developed grammars for the specification of concrete and abstract syntax including parser and AST classes.

In Fig. 9 grammar inheritance is used to parse Java with a new form of expressions namely SQL select statements (we qualify nonterminals using package names as discussed below). The new production SQLSelect overrides the inherited production SQLSelect from the SQL-grammar by adding a new interface Expression which originates from the Java-grammar. First, this establishes an inheritance relationship between mc.sql.SQLSelect and mc.javasql.SQLSelect. Second, it enforces a subtyping relationship between mc.java.Expression and mc.sql.SQLSelect defined in line 6. In this special case the body of the production remains unchanged because it is not further specified (which is different from an epsilon production). The subgrammar also inherits all token classes from their supergrammars. This allows a developer to override the definition of a token class and thereby to use a different lexical analysis in the subgrammar.

As shown in Fig. 9, each new nonterminal results in a new class in the abstract syntax. MontiCore ensures that this class is a subtype of the class that is defined by the overridden rule and therefore all other classes that refer to original class (for example through compositions or associations) can remain unchanged. This approach is much better than a complete regeneration of new AST classes for the subgrammar, because algorithms that work on the abstract syntax of the supergrammar can be applied to the classes of the subgrammar. This is extremely helpful if complicated algorithms for the extended language, e.g., for symbol table building can be still used (maybe after minor adaptation) but no reimplementation is necessary.

A well-known problem of multiple inheritance is name clashes when different supergrammars use the same production names. See for example [60] for a discussion on the possibilities to resolve these problems for object-oriented programming languages. For the MontiCore grammar format we decided to use the following solution: in the case that two or more supergrammars share a common production name, the new production must be a subtype and thus contain all elements of all productions with that name in the supergrammars. Of course, this is not always possible because the super-productions may contradict each other. But, there are cases where it is possible to override the rule:

1. All equal-named productions of the supergrammar have the same type (which is likely if the supergrammars have common ancestors).
2. One class is already a subclass of all other involved classes.

We decided to realize the above-described solution to avoid explicit resolving strategies like in C++ where the developer can refer to a specific superclass by naming it. We felt that theses references would complicate the grammar too far which contradicts the aim to provide a readable specification of languages. Therefore, we advocate an agile way of developing domain-specific languages, where a refactoring of one of the contradicting supergrammars solves the problem, and the readability of the resulting grammar can be retained.

Methodically, it is desirable to apply grammar inheritance only if the desired language extension is similar to the super-language. For example, in the given Java/SQL example the SQL productions add new keywords like SELECT and FROM to the language which are no longer valid identifiers for the new language. This situation and the consequences are similar to the introduction of the assert keyword (and necessary subsequent changes to legacy software) in Java 1.4. If this situation implies serious problems or the two languages contradict each other in form of the lexical analysis, language embedding should be used that allows using separate lexers and parsers for the two languages. Furthermore, language embedding decouples not only the language definition, but also parsers and tooling infrastructure.

### 3.2 Language embedding

DSLs are usually designed to solve a clearly defined task; therefore, it is often necessary to combine several languages in order to define all aspects of the artifact of interest. A typical representative for language that needs to be combined is OCL. It is able to work as a constraint language for other models and must therefore be combined with another language in order to define the artifact of interest more precisely.

For an integrated management, it is convenient to write an OCL statement nearby the artifact that it is constraining. Therefore, OCL statements shall be embedded in another given language. Using standard parsing technologies this would require a single grammar containing all involved sublanguages, which hinders the reuse of single sublanguages and results in monolithic grammars. Instead, we prefer different languages that can be flexibly combined with each other.

The MontiCore grammar allows defining external nonterminals. These are nonterminals where other languages can be hooked into in order to continue parsing according to their grammar. Figure 10 gives an example for language embedding. The keyword external marks the nonterminals StatementCredit and StatementCash as external which have to be filled with appropriate productions in the embedded language. In addition, the nonterminal StatementCash specifies the constraint that the start rule of the embedded grammar must return an instance of the type example.IStatementCash. The slash marks it as a handwritten interface; therefore, it is possible to define requirements for the embedded language in the form of methods that have to be supplied. It is noteworthy that no specific language is embedded here, but the type of the top node of the embedded language is specified.

External nonterminals can be understood as "bottom nonterminals" in grammar fragments [42]: they can be used on the right-hand side of a production but are not defined as production in the grammar itself.

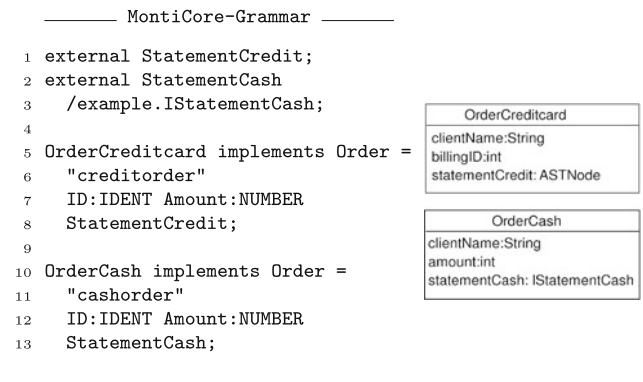

```
                    MontiCore-Grammar
1  external StatementCredit;
2  external StatementCash
3    /example.IStatementCash;
4
5  OrderCreditcard implements Order =
6    "creditorder"
7    ID:IDENT Amount:NUMBER
8    StatementCredit;
9
10 OrderCash implements Order =
11   "cashorder"
12   ID:IDENT Amount:NUMBER
13   StatementCash;
```

**Fig. 10** Language embedding through external nonterminals

The MontiCore tooling infrastructure is able to independently derive parsers and abstract syntax from these grammars and to combine those parsers at runtime. MontiCore ensures the correct behavior of the underlying parsing algorithms: it invokes the embedded parser in order to recognize the subsequent text according to the embedded language.

The most complex situation in this setting is when more than one language is used to replace one single external nonterminal. MontiCore can make the decision within the base language as it is able to select the correct language by values of already parsed attributes or by using predicates. For this case, we add an additional attribute lang:IDENT to the rule OrderCash in Fig. 10. After this has been done, MontiCore can be instructed to use Java whenever the value of the attribute is "Java" and C++ otherwise. This approach is recommended when the embedded languages interfere with each other, i.e., when there are sentences which are valid for both Java and C++. The introduction of an explicit attribute allows the user to select the correct language himself.

The resulting grammar can be seen as the union of all involved fragments. However, and this is the most important advantage, languages can be developed and embedded independently of each other. Furthermore, as explained in more detail in Sect. 4, analysis and generation tools can be developed in the same modular fashion. In addition to the parsers and the AST also the processing algorithms are decoupled.

Our experiences with language embedding show that often the designers of the host language know that an extension is needed, but the form of extension is unclear at design time. This distinguishes language embedding from language inheritance where at design time it is usually unknown that an extension is needed and at a later point in time the existing language shall be altered. The Statechart language, for example, supports actions on the transition, but the exact language in which actions are specified is left open and may change according to the operational environment. Therefore, it is natural for the language designer to specify a "hole" in the grammar in form of an external rule Action where different languages can be plugged in. Since a Java grammar is included in the MontiCore framework, it is most convenient to use it as a default action language in such cases as shown in [22]. However, there are often situations where a combination of embedding and inheritance leads to desired results. As both concepts do not interfere with each other, MontiCore supports to use them in parallel.

## 4 Developing tools in a modular fashion using the DSLTool-framework

The MontiCore grammar defines parsing components that transform the linear text of a DSL into an object structure in modular fashion. This enables reuse and is a prerequisite for libraries that contain language definitions for reuse. A parser, however, only forms the frontend of the language processing framework. The consecutive steps like analysis and code generation also have to be designed in a modular fashion to gain the full benefit from such an approach. MontiCore provides the DSLTool-framework that is designed to support an easy realization of generative and analytic tools that operate on the DSL's abstract syntax. In the following we describe the features of the DSLTool-framework, its architecture, and a subset of its features in more detail. We especially focus on the compositional aspects of language engineering.

4.1 Architectural drivers and main features

The architecture of such a framework is different from the architecture of a compiler, because model-based code generators often support more than a single code generation like the creation of production, simulation, and test code. Also a commonly accepted intermediate language for different kind of models is not established. Therefore, we did not adapt an existing compiler architecture but identified reoccurring technical questions and provided proved and tested solutions for them. The DSLTool-Framework combines these solutions as a basis for specific generative tools. We identified the following architectural drivers [1] for such a framework:

1. Modular decoupled development of algorithms for a given language.
2. Integration of different languages and algorithms in the same tool.
3. Flexible configuration.
4. Easy-to-use APIs for reoccurring tasks within generative software development.
5. Executable on different platforms.

From this list of architectural drivers a set of functions was derived that are supported by the DSLTool-Framework. The creation of various tools for DSLs and especially the generator for the MontiCore grammar format in the bootstrapping process was helpful to get feedback on the framework design.

- Attributes. Attribute grammars [35] are ways to specifiy attributes that are calculated according to the data in the AST. MontiCore grammars can be enriched by adding attribute definitions. Attribute calculations are defined in Java. Details can be found in Sect. 4.6.
- Error messages. Understandable error messages are an important aspect of language development. Generative tools should show descriptive error messages for faulty inputs to users, so that the input can be corrected easily. The error message implementation does not depend on the execution environment and provides means to add new ways of error reporting (for example in an Eclipse plugin).
- File creation. A standardized and simple way to create files and folders supports developers to concentrate on their actual task. Therefore, the DSLTool-Framework offers means to easily create files or folders. Furthermore, within the file creation of a generative tool it is important not to write the same file repeatedly and switch off file creation completely for test cases. The first approach increases the performance of coupled tools like compilers, and the second approach ensures that test cases are free of side effects. Using the DSLTool-Framework nonrecurring file creation is automatically ensured, switching off file creation can be archived in the configuration.
- Functional API. Manipulation of data can often be described in a concise functional form. Therefore, the developer is supported by a Java API with functions as first classartifacts.
- Incremental code generation. The creation of output files whose content is based on multiple input files hinders compilation of just the modified input. The DSLTool supports an effective intermediate storage of partial files to enable incremental code generations and therefore speeding up the generation process.
- Model management. The processing of different models which refer to each other makes it necessary to have a name system between different types of models. The DSLTool-Framework ensure the principle interoperability between different models as explained in further detail in Sect. 4.3.
- Order of processing. The order of the steps within the processing of models should be parameterizable. Thus, a tool can be created that shows a distinct behavior depending on its runtime configuration.
- Platform inpendence. Generators based on the DSLTool-framework are executable as command line tools. This simplifies the integration in continuous build systems. Additional plugins can be generated that integrate the tools in Eclipse without changing the generator logic. Details can be found in Sect. 4.5.

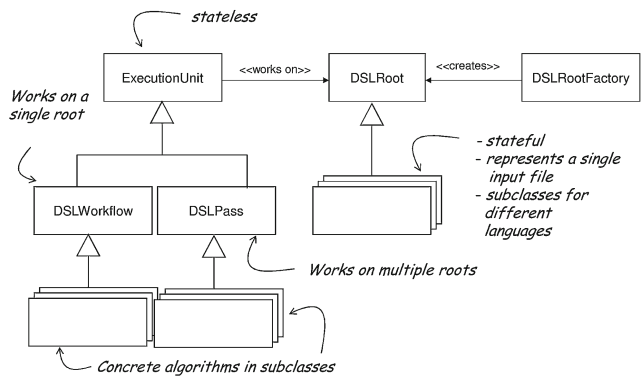

**Fig. 11** Relationship between Roots, ExecutionUnits and RootFactories

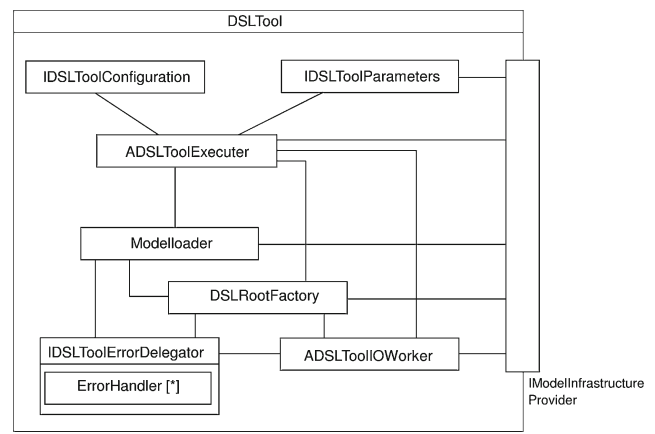

**Fig. 12** Overview of a DSLTool

Template-Engine Generators can be written with popular template engines like Velocity [64] or Xpand [53]. In addition, the DSLTool-Framework provides its own template engine that supports co-evolution of templates and runtime environments which simplifies the agile development of generators including Refactoring [37].

Traversal of data structures. Flexible traversal strategies for data structures are needed within generative tools. The solution for the DSLTool-Framework is described in detail in Sect. 4.4.

### 4.2 Architecture

The DSLTool-Framework processes object structures that contain the data of input files in a structured form. Depending on the file format of the input files these objects have a varying type and are called root objects. The common supertype of all root objects is DSLRoot which contains basic functionality for accessing the generator, file processing, and status/error message. In addition, it contains information like the AST and symbol tables and acts as a repository to store additional information computed by execution units which is needed for subsequent computations.

Root objects are created by a RootFactory which sets up parsers and pretty printers. Standard subclasses exist to distinguish different languages by the used file extension or the first used keyword in the instance. An Execution-Unit encapsulates algorithms that operate on ASTs, beginning with the AST's root object. The algorithm is assumed to be stateless. Instead, all calculated state information shall be placed in the AST or directly on the root object as annotations. Subclasses exist to process single root objects or sets of root objects at the same time. Figure 11 summarizes the relationships between these most important classes in the life cycle of a root object.

A generator within the DSLTool-Framework is a subclass of DSLTool. Its internal structure is displayed in Fig. 12. Its configuration (IDSLToolConfiguration) includes the root object types and available algorithms. A subset (or all of them) will be executed on a given input depending on the runtime parametrization stored in IDSLToolParameters.

The order of processing is determined by the class ADSL-ToolExecuter. During the execution it can be necessary to load additional models from the file system which is be done by the Modelloader in collaboration with the DSLRootFactory as explained above because these additional models are also represented as root objects. Files are read and written by ADSLToolIOWorker which simplifies testing. Errors and status messages are processed by the IDSLToolErrorDelegator that feeds different Errorhandlers for different platforms like command line and Eclipse. The DSLTool can be accessed from algorithms realized as ExecutionUnits by the interface IModel-InfrastructureProvider which provides a subset of the functionality of a DSLTool to the developer.

### 4.3 Model management

Packaging is a well-known concept to organize projects and thus, to handle complexity. MontiCore offers standard infrastructure to add a package mechanism to a DSL definition that is similar to Java packages. If the DSL definition incorporates the provided standard form of packaging, DSL writers are able to use qualified names that consist of a package name (dot-separated identifiers) and the name of the DSL instance (model) itself. In those DSL instances it is possible to optionally start with a declaration to which package they belong. To support a reasonable project structure, the conformance of the package declaration with the file system is checked automatically: models in the package A.B.C must be located in a directory A/B/C.

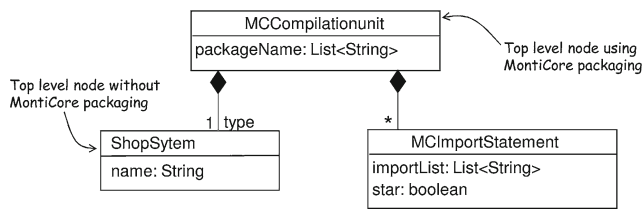

**Fig. 13** Excerpt of abstract syntax using MontiCore packaging

In the implementation the name of the instance itself has to be provided by the DSL, the top level AST-node must return the name of the instance by a `getName()`-method. This method can either be handwritten or generated; the latter is the case if the corresponding rule has an attribute `name`. As for package names, the conformance of the DSL instance name and the filename is checked automatically. For a completion of the package structure, it is possible to add a list of imports behind the package declaration with the same structure and effect as in Java. Imports can be used in order to load and resolve instances of other packages. This functionality is already supported by the MontiCore framework and therefore easy to use in a new DSL definition.

As said, the usage of the MontiCore package mechanism is optional. It is added to the DSL by adding the `compilationunit` keyword in the options section of the grammar followed by the start rule which is responsible for providing the name of the instance. The usage of packaging results in a slight different abstract syntax as outlined in Fig. 13.

The methods of the interface ModelInfrastructureProvider to load models is based on the implementation of the `getName()`-method and the use of the option `compilationunit`.

### 4.4 Visitor

The traversal of object structures is an often needed auxiliary tool to implement analyses and code generations. MontiCore supports the traversal of models along their spanning composition tree and uses a adapted Visitor design pattern [19]. The realization of the traversal is based on Java Reflection to support the dynamic extensibility provided by language embedding. The central class for the traversal is `mc.ast.Visitor` which can be parameterized with different `mc.ast.ConcreteVisitor` instances for different fragments. This fits to the embedding of languages where algorithms should be implemented for the fragments so that they be combined at configuration time to form a complete algorithm for the composed language. Language inheritance is supported by subtyping the `ConcreteVisitor`.

Figure 14 shows a combination of two `ConcreteVisitor` objects to a single `Visitor`. The Visitor traverses an object structure and fails if either a `CashOrder`

```Java
1  public class Test extends DSLWorkflow<ShopSystemRoot>{
2
3    // ...
4    @Override public void run(ShopSystemRoot dslroot) {
5
6      ASTShopSystem ast = dslroot.getAst();
7
8      Visitor v = new Visitor();
9      v.addClient(new CheckNoCashOrders(x));
10     v.addClient(new CheckBalance(x));
11
12     boolean success = v.transform(ast);
13   }
14 }
15 public class CheckNoCashOrders extends ConcreteVisitor{
16
17   public void visit(OrderCash x){
18     getVisitor().fail();
19   }
20 }
21
22 public class CheckBalance extends ConcreteVisitor{
23
24   public void visit(OrderCreditCard x){
25     if (x.getAmount<10){
26       getVisitor().fail();
27     }
28   }
29 }
```

**Fig. 14** Example for compositional visitors

or an `OrderCreditCard` with less then 10 Euro is used. The first part of the functionality is realized on the shop system language, whereas the credit card order is realized on a fictitious language embedded in the external nonterminals shown in Fig. 10.

The traversal of the object structure is a preorder run along the spanning composition tree of the model. The run can be influenced by the developer using the following three types of methods which all have a single parameter that has the type of a model class. An arbitrary mixture for different model classes is possible:

visit(...)      This method is invoked before the child nodes are traversed.
endVisit(...)   This method is invoked after the child nodes are traversed.
ownVisit(...)   This method is invoked before the child nodes are traversed and stops further traversal for these children.

In addition, within the visit-methods the developer can invoke the method `stopTraverseChildren()` to stop the traversal of the children so that it behaves like a `ownVisit(...)`-method. As an alternative the traversal can be stopped at all by invoking `stopTraverse()`. Conversely, it is possible to explicitly start the traversal of child nodes by invoking the `startTraverse(...)`-method on them.

In contrast to other approaches [8,54] we stick to the basic pattern from [19] and generate a `traverse(...)`-Method within the model classes. This invasive version speeds up the execution for the pure traversal: other non-invasive approaches showed a performance reduction of factor 18 to 256 compared with the basic implementation; we could only measure a slow down of factor 4. The time for traversal is usually only responsible for part of the runtime. For more complex operations like a code generation we could therefore measure a total overhead of about 30–40% in comparison with using the basic pattern for the same algorithm. As the implementation allows intentionally to not traverse certain subtrees by using `ownVisit(...)`-methods, our implementation can also be quicker depending on the given algorithm.

The combination of different Visitors is possible as all Visitors can fail (as already shown in the example). Therefore, a combination of Visitors to strategies as explained in detail in [43,66,67] is possible and realized within the MontiCore framework.

### 4.5 Eclipse

Being a sophisticated editor generator is not the main focus of MontiCore but an exploration of concepts for DSL definition and handling. However, the usability of a language highly depends on how the tools support the users when working with the language. Nowadays, there are specialized editors and IDEs like Eclipse [12] for almost every language with comfort functions like syntax highlighting or auto-completion. Even in Eclipse a modification of the underlying language usually results in time-consuming modifications for language-specific tools. This is an obstacle for agile language development and evolution. Therefore, it is desirable to combine language and tool development in an efficient way.

For effective language-specific tool development MontiCore offers possibilities to generate Eclipse plugins. A MontiCore grammar can be complemented with a small editor description that supports customization of an DSL-specific editor similarly to the IMP [61] approach.

As a side note, it is noteworthy that we use an entirely different technique. In contrast to the code-centric approach of IMP DSL developers do not have to implement generated skeleton classes. Instead all information necessary for tool definition is integrated into the language specification. Especially, wizards were avoided as they tend to hinder the evolution of a language due to limited round-trip facilities. Beyond that, the tool and the language definition are co-located and thus it is relatively like that they remain consistent.

The most important options are introduced in the following summary: Fig. 15 shows an excerpt from the definition of the plugin for the shopsystem DSL. The generated editor and some of its functionalities can be seen in Fig. 16.

```
------------------- MontiCore-Grammar -------------------
1  keywords: creditorder, cashorder, client,
2           premiumclient;
3
4  foldable: ShopSystem, PremiumClient, Client;
5
6  segment: ShopSystem ("pict/mc.ico") show: name;
7  segment: PremiumClient ("pict/premium.gif")
8           show: name;
9  segment: Client ("pict/client.gif") show: name;
10 segment: OrderCreditcard ("pict/credit.png")
11          show: clientName ;
12 segment: OrderCash ("pict/cash.gif")
13          show: clientName;
14
15 editoritem Generate Code ("examples.CodegenAction");
16
17 menuitem Merge shops ("examples.MergeAction");
```

**Fig. 15** Excerpt from a definition of a language-specific editor

- Syntaxhighlighting. Syntaxhighlighting is very helpful to easily get an overview of a language document. Language-specific keywords are defined by a comma-separated list and are automatically detected and colored in the generated editor.
- Outline. An outline provides an overview of the language instance in a separate view. A segment in the outline consists of a small icon and a text which can also depend on the attribute values of the AST node it represents. Selecting an item in the generated outline marks the representing code area within the editor.
- Folding. Folding provides functionality to show and hide parts of the language document. Nonterminals that should provide folding functionality can be defined in a comma separated list.
- Error messages. Error messages containing a declarative description of the error and the area of the erroneous language part are shown in the problems view of Eclipse. Errors typically occur while the text is parsed or during the check of contextual constraints. It is possible to hook in self-written checks. Again, selecting an item in the problems view marks the according code in the text editor.
- Editor menu items. User-defined functionalities can be hooked into the generated plugin by menu items like "transformations" or "code generation" in the context menu of the generated editor.
- Explorer menu items. Similarly, it is possible to define popup menu items for the context menu of the package explorer (a view which shows whole projects and all files, etc. within the project). They allow to hook in functionality which depends on more than one file (which is mostly the case for editor menu items). A composition of several models is an example for explorer menu items.

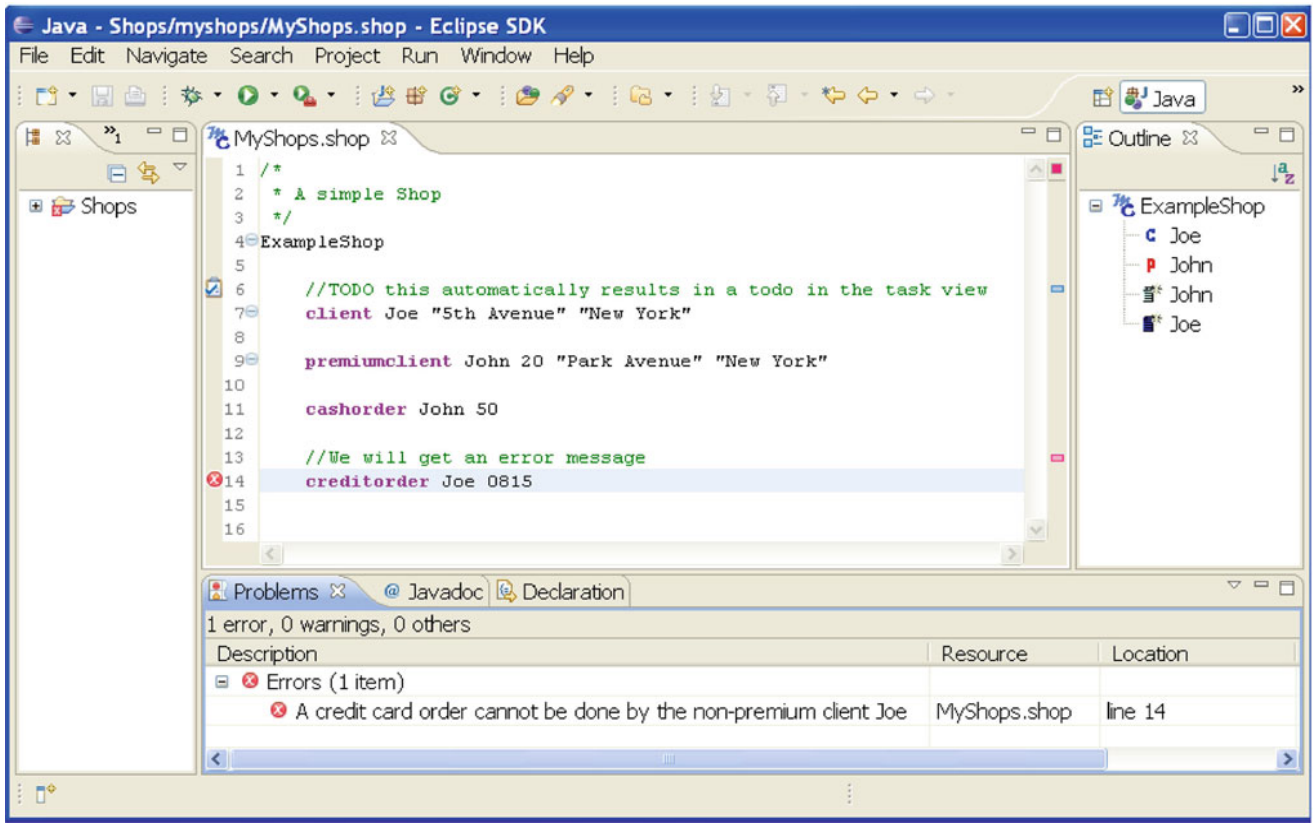

**Fig. 16** Screenshot of the generated editor

Modularity. All concepts described above support modularity. This means that a combination of two languages by embedding automatically leads to new tools that support the combined functionality for this combined language. Furthermore, in the case of inheritance it is sufficient to specify the delta for the sublanguage only and more important, only the delta of the editor functionality is generated.

4.6 Attributes

MontiCore offers a flexible attribute grammar system which reflects the compositional approach for the development of abstract and concrete syntax of a DSL as described above. Each grammar can be supplemented by an arbitrary number of attributes which are either synthesized or inherited (c.f. [35]). As an example for the shopsystem we introduce an attribute `outstanding` which is used in order to compute the outstanding accounts for the whole shop. Therefore, we augment our grammar with the definition of a synthesized attribute as shown in Fig. 17.

This definition is not complete as the actual computation is missing. In order to compute values, we decided not to use a specialized attribute computation language but to use Java. This enables the developer to program complex compu-

```
─────────────────────── MontiCore-Grammar ───
1 // synthesized  attribute "outstanding".
2 // The type of the attribute is float.
3 syn outstanding: /float
─────────────────────────────────────────────
```

**Fig. 17** Example for a synthesized attribute definition

tations as well as to use an arbitrary type for each attribute. The attribute calculations are implemented in a Java class as methods which compute the values as shown in Fig. 18. Line 22 requires a detailed discussion: As explained earlier, MontiCore grammars can be combined flexible using grammar inheritance and embedding. One can imagine that in the case of embedding the outer language has an attribute `outstanding`, whereas the inner language has an attribute `sum` which are semantically the same. When combining both languages we wanted to avoid that one of the calculations has to be adapted.

To solve this issues we used the following strategy: each grammar defines its attributes as explained above. Then, when combining languages the developer can map different attributes of different languages to one "virtual" attribute. This can be done either by a special DSL (see Fig. 19) or by writing some glue code in Java. We prefer these "virtual" attributes

```Java
public class OutstandingCalculation
    extends SynthesizedAttributeCalculation {

    // for each class of the abstract syntax
    // a calc-method is implemented which does
    // the actual computation
    public float calc(OrderCash c){
        int discount=
                getDiscount(c.getOrderingClient());
        return c.getAmount() * ( 1 - discount/100);
    }

    public float calc(OrderCreditCard c){
        int discount=
                getDiscount(c.getOrderingClient());
        return c.getAmount() * ( 1 - discount/100);
    }

    public float calc(ShopSystem s){
        float f=0;
        for (Order order:getOrders()){
            f+=getConnector().getOutstanding(order);
        }
        return f;
    }

    // just a helper to compute the discount for
    // a client depending on the type of the client
    protected int getDiscount(Client c){
        if (c instanceof PremiumClient){
            return ((PremiumClient)c).getDiscount();
        }
        return 0;
    }
}
```

**Fig. 18** Example for a synthesized attribute calculation

```MontiCore-Grammar
//introduce a virtual attribute "Unpaid"
Unpaid{

    // The attribute "outstanding" of the grammar
    // mc.examples.shopsystem is calculated in
    // mc.examples.shopsystem.OutstandingCalculation.
    // This attribute contributes to the
    // calculation of "Unpaid"
    mc.examples.shopsystem.outstanding =
       /mc.examples.shopsystem.OutstandingCalculation;

    // The same holds for this attribute
    mc.examples.shop2.sum =
       /mc.examples.shop2.SumCalculation;
}
```

**Fig. 19** Example for an attribute combination

over adding new computation rules that map the different attributes for two reasons. First, language embedding might occur on several places in a grammar where this mapping has to be repeated. Second, the combination of fragments with the "virtual" attribute might be reused and embedded in other fragments. Using our approach, it is then not necessary to understand the internal structure of the combination with different attributes, but the user can rely on a single attribute.

## 5 Related work

**Language workbenches**. A language workbench simplifies the development of domain-specific languages by providing formalisms to define the language. There are graphical approaches (e.g., [45]), but we concentrate on approaches that allow the specification of textual domain-specific languages.

The Meta Programming System (MPS) allows the development of textual languages as an extension to an IDE for Java. A syntax-directed editor is generated from the language definition, and a template engine helps the developer to specify code generations. Attribute grammars and their tool suites like LISA [48] allow a grammar-based development of domain-specific languages. In the general sense, a lot of concepts like MontiCore's associations can be realized as attributes in such grammars. MontiCore, furthermore simplifies a number of standard cases by supplying standard solutions that be easily applied by a developer.

The Grammar Deployment Kit (GDK) [36] consists of several components to support the development of grammars and language processing tools. The internal grammar format can be transformed into inputs of different parser generators, such as btyacc [10], Antlr [57] or SDF [28]. Furthermore, it provides possibilities for grammar adaption, like renaming of rules or adding alternatives. In opposition to our approach it does not support extensions like inheritance or associations.

ASF+SDF [2] is a language development meta-environment. The syntax definition is based on a scannerless generalized LR parsing technique [65] and permits modular syntax definition. Furthermore, the framework offers support for source code analysis, transformations, code generation, and IDE development. The main difference to MontiCore is that we offer modularity concepts not only at the syntax level but we reflect these concepts at the level of other aspects of language development like code generation by visitors, attribute grammars, and editor generation.

**Languages and tools for specifying concrete and abstract syntax**. We are currently not aware of a language that allows specifying both a textual concrete syntax and an abstract syntax with additional cross-AST associations in a coherent and concise format. Grammarbased approaches usually lack a strongly typed internal representation (for exceptions see below) and the existing model-based approaches use two forms of description, a meta-model for the abstract syntax and a specific notation for the concrete syntax.

SableCC [17] is a parser-generator that can generate strongly typed abstract syntax trees and tree-walkers.

The grammar format contains actions to influence the automatic derivation of the AST. In contrast to MontiCore, SableCC does not aim to include associations in its AST.

The algorithm presented in [68] derives a strongly typed abstract syntax from a BNF-like grammar. In contrast to the MontiCore parsing frontend Wile uses an explicit notation for lists that are separated by constants and the missing integration of nonterminals with same name.

In [32] a DSL named Textual Concrete Syntax (TCS) is described that specifies the textual concrete syntax for an abstract syntax given as a meta-model. Different meta-modeling techniques can be used with the approach like KM3 [31] or EMF [7]. The described tool support is similar to the one we used for the MontiCore framework and the name resolution mechanisms are the same that we generate automatically from the grammar format. In contrast to our approach, two descriptions for abstract and concrete syntax are needed.

In [15] and [50] the Textual Concrete Syntax Specification Language (TCSSL) is described that allows the description of a textual concrete syntax for a given abstract syntax in form of a meta-model. TCSSL describes a bidirectional mapping between models and their textual representation. The authors describe tool support to transform a textual representation to a model and back again. As in MontiCore the AST usually (but not necessarily) is a real abstraction the AST loses the necessary information to keep bidirectional links. However, we are more interested in transformation, AST-based analysis, and code generation and therefore need not retain the original concrete syntax in those cases.

**Compositional language development**. Compositional language development is an important goal, especially in grammar-based software. The main problem is that common techniques as LL or LR parsing are not closed under composition. A particular problem using LL or LR is on the lexical level, [6] discusses different solutions, one of them—namely controlling the lexer from the parser—is implemented in our tool although [6] favors another strategy (scannerless parsing) mainly for technical reasons. The strategy of controlling the lexer from the parser is also used in [69]. In opposition to our approach, a parser controls one single lexer by passing all tokens that are possible at this point of parsing to the lexer. We generate different lexers for different languages which are selected at runtime. Therefore, we can reuse parsers and lexers without regeneration/recompilation. The same approach (passing valid tokens to the lexer) is implemented in the Silver system [63]. Both Silver and MontiCore permit multiple language inheritance and thus, the combination of multiple languages, but the approaches are slightly different. While Silver combines languages and generates parser/lexer from the combined version, we keep the languages standalone and combine them at configuration time. Therefore, we do not need to regenerate everything when only one language changes. This seems to be more appropriate because all sublanguages are often not known. Furthermore, Silver offers an attribute system and forwarding techniques [70] to implement language extensions. This attribute system uses a special DSL to express computations and forwarding while we use Java.

In addition, there are sophisticated parsing technologies which permit a compositional approach (e.g., GLR [62], Early parsers [11], or Packrat parsing [16,21]. These technologies often concentrate on the concrete syntax only, whereas our approach integrates all parts of language development in a compositional manner.

Beyond these parsing technologies, attribute grammar systems exist that permit a modular language development. Well known examples are the LISA [47,48] and JastAdd [13,14]. Especially, JastAdd provides a lot of support for different kind of attributes, amongst them Reference Attribute Grammars (RAGs) [27]. RAGs permit attributes to be references to nodes in the AST. This is comparable to our association concept as in both MontiCore and JastAdd users define the rules for attribute computation in Java. The main difference to our work is that JastAdd mainly concentrates on the specification of attributes and extensibility of compilers. This requires the developer to use other tools, e.g. for parser generation. MontiCore provides an integrated solution with a uniform frontend by embedding external tools in a transparent way.

Language libraries as discussed in [6] mainly target at GPLs with embedded DSLs. These DSLs are the assimilated into the host language to design a language extension. A prominent example for this approach is MetaBorg [5,4]. However, we do not concentrate on GPL extensions (although this can be done using MontiCore) but on the co-existence of several languages on the same level. This seems to be more appropriate in the DSL world as we usually have several languages specialized for a specific task and thus, there is often no possibility to map one language into another.

Polyglot [51] is an extensible compiler framework for Java. It provides an infrastructure to implement extensions on the level of concrete syntax, abstract syntax, type system, and code generation. These extensions are implemented using object-oriented methods like inheritance, delegation, and factories. In this respect it is comparable to the principles used in our framework MontiCore. However, Polyglot it is limited to Java and extensions thereof. Although our framework can be used for the same purpose, we support arbitrary languages.

## 6 Conclusion

As main results, this work discusses the possibilities of modular, compositional language development in MontiCore, and how embedding of languages and language inheritance can be achieved.

MontiCore is text-based and uses an extended grammar format to specify both, abstract and concrete syntax of a modeling language in a concise format. By using an integrated representation for both it avoids typical redundancy problems that occur when abstract and concrete syntax are described by two different languages. To generalize from tree structures to graphs (with spanning trees), the possibility to define associations between AST nodes and provide standard functionality to establish those links after parsing through name resolution techniques was added.

MontiCore provides two modularity mechanisms to reuse existing languages in a controlled way. First, grammar inheritance allows extending a grammar A in a subgrammar B by extending the nonterminals from A with new parsing alternatives. This allows keeping A and its generated code unchanged and therefore paves the way for extensible languages. Language inheritance allows subtyping a language in order to adapt it to new needs. Second, language embedding allows specifying a grammar A(h) with explicit holes h by identifying one or more nonterminals that are not realized within the language definition itself. Another language B is embedded into A(B) by filling the hole with an appropriate nonterminal. While this is theoretically relatively straightforward, the MontiCore framework can generate code for the parsers as well as symbol tables and other infrastructure independently and compose the parsers at configuration time. This is a very important new feature and paves the way for a modular language definition and even reuse of infrastructure when the source code is not available.

As said our techniques are implemented in a framework called MontiCore, which is based on an established parser-generator. It is able to parse textual syntax and generates the model representation in Java. Additionally, EMF support is available to be interoperability with a variety of other tools. We have used the framework to develop tools for a number of toy examples as well as sophisticated language definitions like UML/P [58,59] and complete Java 5. In addition, the system is bootstrapped and currently about 75% of the code is generated from several DSLs. The Monticore framework can be used as an online service that is available via [49].

**Acknowledgments** The work presented in this paper is partly undertaken as a part of the MODELPLEX project. MODELPLEX is a project co-funded by the European Commission under the "Information Society Technologies" Sixth Framework Programme (2002–2006). Information included in this document reflects only the authors' views. The European Community is not liable for any use that may be made of the information contained herein.